\documentclass[aps,pra,superscriptaddress,twocolumn]{revtex4-1}

\usepackage{mathrsfs}
\usepackage{amsmath}
\usepackage{amssymb}
\usepackage{amstext}
\usepackage{graphicx}

\begin{document}

\title{Fidelity of the surface code in the presence of a bosonic bath}

\author{P. Jouzdani}

\affiliation{Department of Physics, University of Central Florida,
  Orlando, Florida 32816-2385, USA}

\author{E. Novais}

\affiliation{Centro de Ci\^encias Naturais e Humanas, Universidade
  Federal do ABC, Santo Andr\'e, SP, Brazil}

\author{E. R. Mucciolo}

\affiliation{Department of Physics, University of Central Florida,
  Orlando, Florida 32816-2385, USA}

\date{\today}

\begin{abstract}
We study the resilience of the surface code to decoherence caused by
the presence of a bosonic bath. This approach allows us to go beyond
the standard stochastic error model commonly used to quantify
decoherence and error threshold probabilities in this system. The full
quantum mechanical system-bath dynamics is computed exactly over one
quantum error correction cycle. Since all physical qubits interact
with the bath, space-time correlations between errors are taken into
account. We compute the fidelity of the surface code as a function of
the quantum error correction time. The calculation allows us to map
the problem onto an Ising-like statistical spin model with two-body
interactions and a fictitious temperature which is related to the
inverse bath coupling constant. The model departs from the usual Ising
model in the sense that interactions can be long ranged and can
involve complex exchange couplings; in addition, the number of allowed
configurations is restricted by the syndrome extraction. Using
analytical estimates and numerical calculations, we argue that, in the
limit of an infinite number of physical qubits, the spin model sustain
a phase transition which can be associated to the existence of an
error threshold in the surface code. An estimate of the transition
point is given for the case of nearest-neighbor interactions.
\end{abstract}


\maketitle

\section{Introduction}
\label{sec:introduction}

Recent progress in implementing controllable multiqubit systems in the
laboratory has sparked renewed interest in topological quantum
computing schemes. Particular attention has been devolted to the
surface code \cite{bravyi1998,freedman2001,dennis2002}, which is a
planar version of Kitaev's toric code \cite{kitaev1997}. From a
physical implementation viewpoint, the surface code has two important
advantages in comparison to other schemes: (i) all gates are local,
and (ii) simulations indicate that the topological protection yields
very high tolerance for errors. The latter is based entirely on
stochastic models for errors. These models point to error threshold
probabilities per single qubit operation or cycle ranging from 1\%
\cite{raussendorf2007,wang2011}, when only nearest neighbor
interactions and no perfect gates are assumed, up to 19\%
\cite{bombin2012}, when the ability to perfectly measure four-qubit
operators is assumed. The large error threshold comes at the expense
of hardware: a vast number of local operations and physical qubits is
required to build a useful computing machine \cite{fowler2012}. Yet,
this tradeoff seems attractive nowadays for a number of physical
realizations such as cold atoms \cite{bloch2012}, ion traps
\cite{blatt2008}, Rydberg atoms \cite{saffman2010}, semiconductor
systems \cite{qdotqubits}, and superconducting integrated systems
\cite{scqubits}.

Most error threshold estimates so far have relied on the assumption of
errors being uncorrelated in time and space. However, given the
large-scale integration that will be required to implement a surface
code, this assumption seems unwarrantable on physics grounds. The need
to have tens of millions of physical qubits siting on a common
substrate and interacting with each other and with the controlling
electronics is very likely to introduce environmental modes, which will
effectively couple the time evolution of the physical qubits. Under
these circumstances, errors will become correlated and it is unclear
whether the system will retain its high error threshold. In fact,
previous studies of the impact of correlated errors on standard
(non-topological) quantum computing codes have shown that error
thresholds may be reduced or altogether disappear in some situations
\cite{novais2007,novais2008,novais2010,AKP07,Preskill2009,Preskill2013}.
Investigating the effect of correlations between errors in the surface
code is the main goal of this paper.

In a recent paper \cite{novais2013}, two of us showed that the time
evolution of the surface code in the presence of a common bosonic bath
can be mapped onto a statistical spin model. This mapping allows for
the computation of the surface code fidelity much in the same way that
one computes the partition function and expectation values in a spin
model. As a result, the existence of an error threshold was related to
the existence of a phase transition in the statistical model. Even
though the interpretation of the crossing of the error threshold as a
classical phase transition is not new \cite{aharonov2000,dennis2002},
our formulation takes into account the full quantum mechanical time
evolution of the qubits in the presence of a dynamical environment. In
addition, rather than evaluating error probabilities, we compute
directly the fidelity of the logical qubit. Our choice of environment,
a collection of freely propagating massless bosonic modes, is
realistic for systems where decoherence can be related to the coupling
to phonons, magnons, and electromagnetic modes.

Below, we provide a detailed derivation of the evolution operator of
the combined surface code--bosonic bath system. We focus our attention
on a single quantum error correction cycle and assume that, after the
syndrome extraction, the bath is reset to its ground state. Within
this approximation, we find that the fidelity can be written as a
function of the expectation value of single-qubit logical operators.
The study of these expectation values can be related to the physics of
an Ising-like spin model with a complex fictitious temperature. Under
the assumption of noncyclic and perfect stabilizer measurements, we
use both exact and mean-field finite numerical calculations to argue
that the spin model sustains a thermodynamic phase transition in the
limit of an infinite number of physical qubits. System with 25 and 41
qubits are studied numerically. The critical temperature of the spin
model yields a coupling constant threshold value which is found to
depend mainly on bath parameters.

The paper is organized as follows. In Sec. \ref{sec:surfacecode} we
give a give a brief introduction to the essential elements of the
surface code and set some of the notation used later. Section
\ref{sec:bosonic} presents a Hamiltonian formulation of the problem in
terms of bosonic modes coupled to physical qubits which allows us to
obtain a compact form for the evolution operator of the combined
logical qubit--bath system. The evolution operator involves a bath
correlation function which is explicitly evaluated for three
representative situations. The effect of syndrome extraction on the
evolution operator is described in Sec. \ref{sec:QEC} and an
expression for the fidelity in terms of expectation values involving
qubit operators is derived in Sec. \ref{sec:fidelity}. The mapping of
the fidelity calculation onto a statistical model is given in
Sec. \ref{sec:statmodel} and the connection between the fictitious
critical temperature and the error threshold probability is shown in
Sec. \ref{sec:errorprob}. In Sec. \ref{sec:estimate} we estimate the
fictitious critical temperature via a low-temperature
expansion. Numerical supporting the existence of an error threshold
are shown in Sec. \ref{sec:numerics}, which is a very encouraging
result. Conclusions and a critical discussion of the approximations
involved and future directions of investigation are drawn in
Sec. \ref{sec:conclusions}. A number of appendixes with technical
details of the calculations are also provided.

\section{Surface code}
\label{sec:surfacecode}

\begin{figure}[ht]
\includegraphics[width=0.6\columnwidth]{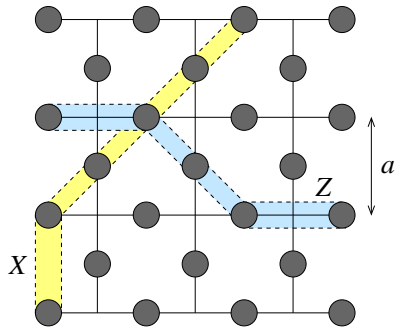}
\caption{(Color online) A $3\times 3$ two-dimensional square lattice
  structure for a surface code. Physical qubits (circles) are located
  at the edges of the lattice, which has open (vertical) and closed
  (horizontal) boundaries. In the general case, the lattice is $n
  \times m$ in size, with $nm$ vertical and $(n+1)(m+1)$ horizontal
  edges and $N = $ physical qubits. The light colored strips show
  possible paths for the logical operators $\bar{Z}$ and
  $\bar{X}$. $a$ is the lattice constant.}
\label{fig:Lattice}
\end{figure}

Following Ref. \cite{bravyi1998}, we define the surface code as a
collection of $N$ spins 1/2 (physical qubit systems) located on the
edges of a two-dimensional lattice with two types of boundaries, as
shown in Fig. \ref{fig:Lattice}. The lattice comprises $n$ and $m$
qubit rows and columns, respectively. Measurements are done on two
types of stabilizer operators: stars $A_{\lozenge}$, which are
associated to lattice vertices ($\lozenge$),
\begin{equation}
\label{eq:A}
A_{\lozenge} = \prod_{i\in \lozenge} \sigma_i^x,
\end{equation}
and plaquettes $B_{\square}$, which are associated to tiles
($\square$), including the ones at the open boundaries,
\begin{equation}
\label{eq:B}
B_{\square} = \prod_{i\in \square} \sigma_i^z.
\end{equation}
In Eqs. (\ref{eq:A}) and (\ref{eq:B}), the Pauli spin operators
$\vec{\sigma}_i$ act on qubits. Thus, there are $N_{\square} = (n+1)m$
plaquette operators and $N_{\lozenge} = (m+1)n$ star operators. The
$N$ physical qubits store one logical qubit. There are $n_L=2$
distinct logical operators: $\bar{X}$ and $\bar{Z}$. They are formed
by a string of physical qubit operators along paths that cut through
the lattice:
\begin{equation}
\bar{Z} = \prod_{i\in \Gamma_Z} \sigma_i^z
\end{equation}
and
\begin{equation}
\bar{X} = \prod_{i\in \Gamma_X} \sigma_i^x
\end{equation}
where $\Gamma_Z$ runs between qubits at opposite open boundaries (left
to right), passing through vertices along the way, while $\Gamma_X$
runs between qubits at opposite closed boundaries (top to bottom),
crossing tiles (see Fig. \ref{fig:Lattice}). Notice that vertices and
tiles form dual lattices.

The protected code space contains two states, $\left\vert
\bar{\uparrow} \right\rangle$ and $\left\vert \bar{\downarrow}
\right\rangle$. Both states are eigenstates of all stabilizer
operators with eigenvalue $+1$. Errors can be inferred by measuring
the stabilizer operators and tracking down which stars or plaquettes
yielded $-1$ values. A decoding procedure is needed to decide which
recovering operation to perform \cite{poulin10,Fowler12}.

The code state can be generated by the action of a product involving
all star operators on the $z$ ferromagnet state, namely,
\begin{equation}
\left\vert \bar{\uparrow} \right\rangle = G \vert F_z \rangle,
\end{equation}
and
\begin{equation}
\left\vert \bar{\downarrow} \right\rangle = G \bar{X} \vert F_z
\rangle,
\end{equation}
where
\begin{equation}
\label{eq:Gop}
G = \frac {1}{\sqrt{2^{N_{\lozenge}}}} \prod_{\lozenge } ( 1 +
A_{\lozenge})
\end{equation}
and
\begin{equation}
|F_z\rangle = \prod_{i=1}^N \vert \uparrow \rangle_{i,z}.
\end{equation}
Notice that the product in Eq. (\ref{eq:Gop}) can be expanded as
\begin{equation}
\label{G}
  \prod_{\lozenge } (1+A_{ \lozenge }) = 1 +
  \sum_{\lozenge}{A_{\lozenge}} + \sum_{\lozenge_1 \neq \lozenge_2}
  A_{\lozenge_1} A_{\lozenge_2} + \ldots .
\end{equation}
The number $2^{N_{\lozenge}}$ appearing in the prefactor of $G$ is the
number of terms appearing in the expansion: $2^{N_{\lozenge}} =
\binom{N_{\lozenge}}{0} + \binom{N_{\lozenge}}{1} + \cdots +
\binom{N_{\lozenge}}{N_{\lozenge}}$.

\section{Bosonic environment}
\label{sec:bosonic}

The most general bath model would allow for both flip and phase errors
to occur. However, only perturbative calculations would be possible in
this general case. Since our goal is to obtain nonperturbative
results, we focus our discussion on flip errors only (it is possible
to rephrase the model to induce only the pure dephasing by a simple
change of basis). We do not explicitly consider correlated errors
introduced by the hardware upon measurement, but rather errors induced
by the interaction between a bath and the qubits during the time span
of a QEC cycle. The Hamiltonian we consider is written as
\begin{equation}
\label{eq:Htotal}
H = H_{0} + V,
\end{equation}
where $H_{0}$ is a free bosonic Hamiltonian,
\begin{equation}
\label{eq:H0}
H_{0} = \sum_{{\bf k}\neq0}\, \omega_{{\bf k}}\, a_{{\bf k}}^{\dagger}
a_{{\bf k}},
\end{equation}
and
\begin{equation}
\label{eq:V}
V = \frac{\lambda}{2} \sum_i f \left({{\bf r}_i} \right) \sigma_i^{x},
\end{equation}
where ${\bf r}_i$ denotes the spatial location of a qubit $i$ and $f$
is a local bosonic operator,
\begin{equation}
\label{eq:ftr}
f\left({\bf r}\right) = \frac{(v/\omega_{0})^{D/2+s}}{L^{D/2}}
\sum_{{\bf k}\neq0} \left|{\bf k} \right|^{s} \left( e^{i{\bf
    k}\cdot{\bf r}} a_{{\bf k}}^{\dagger} + e^{-i{\bf k}\cdot{\bf r}}
a_{{\bf k}}\right).
\end{equation}
Here, $D$ is the bath spatial dimension, $v$ is the bosonic mode
velocity, $\omega_{{\bf k}}=v|{\bf k}|$, and $\omega_{0}$ is a
characteristic frequency of the bath (notice that $f$ is dimensionless
since we adopt units such that $\hbar=1$). The creation and
annihilation operators of the bosonic modes follow the standard
commutation relations, namely, $[ a_{{\bf k}}, a_{{\bf k}'}^{\dagger}]
= \delta_{{\bf k}, {\bf k}'}$ and $[ a_{{\bf k}}, a_{{\bf k}'}] = [
  a_{{\bf k}}^\dagger, a_{{\bf k}'}^{\dagger}]=0$. 

The choice of $s$ depends on the physical nature of the environment
and on which bosonic degree of freedom couples to the qubits. When the
qubits couple directly to the bosonic displacement field, we choose
$s=-1/2$, whereas when they couple to the bosonic current operator, we
choose $s=1/2$ instead. Notice that these two choices allow us to
write the coupling between the qubits and the bosonic environment as a
simple function of the free bosonic field. Hence, they render an
interaction Hamiltonian with commutators that are subluminal, namely,
which are equal to zero outside the boson light cone (see below). But
this is not the general rule. For instance, a model that couples the
environment to two different qubit components would render an
interaction Hamiltonian with non-subluminal commutators (regardless of
our choice for $s$).

This apparent problem comes from the fact that we usually think of
errors in a dynamical sense: they happen in a point in space-time and
create bosons that propagate at the speed of light. However, this is
an incorrect interpretation to the equations we are about to
derive. We will not be considering pulses propagating thought a
medium, but rather looking at allowed normal modes of the bath and how
they relate to different qubit configurations. If there is no
fundamental symmetry reason for their suppression, long-wave length
modes of the bath will in general introduce superluminal effective
interactions. A very simple way to highlight this fact is to rewrite
the bosonic model in a coherent-state basis,
%
$
a_{\bf k} = \tilde{a}_{\bf k} + \alpha,
$
%
where $\tilde{a}_{\bf k}$ and $\tilde{a}_{\bf k}^\dagger$ also obey
standard commutation relations and $\alpha$ in a constant. This
procedure introduces an effective instantaneous interaction between
the qubits as much as the Coulomb gauge introduces the instantaneous
Coulomb interaction in quantum electrodynamics.

For a time interval $\Delta$, the error model comprised by
Eqs. (\ref{eq:Htotal}) -- (\ref{eq:V}) leads to the following
evolution operator in the interaction picture \cite{novais2010}:
\begin{equation}
\label{eq:UDelta}
U(\Delta) = T_{t}\, \exp \left[ -i\frac{\lambda}{2} \int_{0}^{\Delta}
  dt \sum_{i} f \left({\bf r}_i,t\right) \sigma_{i}^{x} \right].
\end{equation}
Combining a Magnus expansion with the Zassenhaus formula (see Appendix
\ref{sec:appendixA}), we arrive at a remarkably simple expression for
the evolution operator,
\begin{widetext}
\begin{equation}
\label{eq:evolution}
U(\Delta) = \chi\, \exp \left[ -\frac{\lambda^{2}}{2} \sum_{i\neq j}
  \Phi_{{\bf r}_i{\bf r_j}} (\Delta) \, \sigma_{i}^{x} \sigma_{j}^{x}
  \right] : \exp \left[ -\frac{i\lambda}{2} \sum_{i}F_{{\bf r}_i}
  (\Delta)\, \sigma_{i}^{x} \right] :,
\end{equation}
\end{widetext}
where
\begin{equation}
\label{eq:Phi_def}
\Phi_{{\bf r}{\bf s}}(\Delta) = \frac{1}{2} \left[ {\cal
    G}^{(R)}_{{\bf r}{\bf s}}(\Delta) + {\cal G}^{(I)}_{{\bf r}{\bf
      s}}(\Delta) \right],
\end{equation}
\begin{equation}
\label{eq:chi}
\chi = \exp \left[ -\frac{\lambda^2}{4} \sum_i \Phi_{{\bf r}_i {\bf
      r}_i}(\Delta) \right],
\end{equation}
and $::$ stands for normal ordering. In Eqs. (\ref{eq:evolution}) and
(\ref{eq:Phi_def}), we have introduced two bath correlation functions,
\begin{eqnarray}
\label{eq:I}
{\cal G}^{(I)}_{{\bf r}{\bf s}}(\Delta) & = & \frac{1}{2}
\int_{0}^{\Delta}dt_{1} \int_{0}^{t_{1}}dt_{2} \left\{
\left[f\left({\bf r},t_{1}\right), f\left({\bf s}, t_{2} \right)
  \right] \right. \nonumber \\ & & \left. +\ \left[f\left({\bf
    s},t_{1}\right), f\left({\bf r}, t_{2} \right) \right] \right\}
\end{eqnarray}
and
\begin{equation}
\label{eq:GR}
{\cal G}^{(R)}_{{\bf r}{\bf s}}(\Delta) = \langle0|F_{{\bf
    r}}(\Delta)\, F_{{\bf s}}(\Delta)|0\rangle,
\end{equation}
and the auxiliary bosonic field
\begin{equation}
\label{eq:F}
F_{{\bf r}}(\Delta) = \int_{0}^{\Delta}dt\, f\left({\bf r},t\right).
\end{equation}
(Notice that $\chi$ is a real number since ${\cal G}^{(I)}_{{\bf
    r}{\bf r}} = 0$.)

Below, we present the functional form of the correlations functions
for two-dimensional bosonic baths ($D=2$). Details of the calculations
are provided in Appendices \ref{sec:appendixB} and
\ref{sec:appendixC}. We consider three representative values of the
power $s$ which appears in the qubit-bath coupling constant dependence
on the bath mode momentum [see Eq. (\ref{eq:ftr})]. These values are
$s=-1/2$, $0$, and $s=1/2$, corresponding to sub-Ohmic, Ohmic, and
super-Ohmic baths, respectively. This classification is standard and
follows from the bath's spectral function frequency dependence at low
frequencies: sublinear (sub-Ohmic), linear (Ohmic), and superlinear
(super-Ohmic) \cite{weiss}.

\subsection{Sub-Ohmic bath}
\label{sec:subohmicbath}

For two-dimensional sub-Ohmic baths ($D=2$ and $s=-1/2$), the bath
correlation function takes a simple closed form. The real part reads
(see Appendix \ref{sec:real_sub})
\begin{eqnarray}
\label{eq:GRsubohmic}
{\cal G}^{(R)}_{{\bf r}{\bf s}}(\Delta) & = & -\frac{|{\bf r}-{\bf
    s}|}{\pi v \omega_{0}} + \frac{\Delta}{2\omega_0} \theta(v\Delta -
|{\bf r} - {\bf s}|) \nonumber \\ & & +\ \frac{\Delta}{\pi\omega_0}
\theta(|{\bf r}-{\bf s}|-v\Delta) \left[ \mbox{arcsin} \left(
  \frac{v\Delta}{|{\bf r}-{\bf s}|} \right) \right. \nonumber \\ & &
  \left. +\ \sqrt{\frac{|{\bf r}-{\bf s}|^2}{(v\Delta)^2} - 1} \right],
\end{eqnarray}
where $\theta(x)$ denotes the Heaviside step function. The imaginary
part reads (see Appendix \ref{sec:imag_sub})
\begin{eqnarray}
\label{eq:GIsubohmic}
{\cal G}^{(I)}_{{\bf r}{\bf s}}(\Delta) & = &
\frac{-i\Delta}{\pi\omega_0}\, \theta (v\Delta - |{\bf r} - {\bf s}|)
\nonumber \\ & & \times \left\{ \ln \left[
  \sqrt{\frac{(v\Delta)^2}{|{\bf r} - {\bf s}|^2} - 1} +
  \frac{v\Delta}{|{\bf r} - {\bf s}|} \right] \right. \nonumber \\ & &
\left. -\ \sqrt{1 - \frac{|{\bf r} - {\bf s}|^2}{(v\Delta)^2}}
\right\}.
\end{eqnarray}
%

\subsection{Ohmic bath}
\label{sec:ohmicbath}

Choosing $D=2$ and $s=0$, the real and imaginary parts of the bath
correlation function take the forms \cite{errata} (see Appendixes
\ref{sec:real_ohmic} and \ref{sec:imag_ohmic})
\begin{equation}
\label{eq:GRohmic}
{\cal G}^{(R)}_{{\bf r}{\bf s}}(\Delta) = \frac{1}{\pi\omega_{0}^{2}}\,
\mbox{arcosh} \left( \frac{v\Delta}{|{\bf r}-{\bf s}|} \right)\,
\theta (v\Delta - |{\bf r}-{\bf s}|),
\end{equation}
and
\begin{eqnarray}
\label{eq:GIohmic}
{\cal G}^{(I)}_{{\bf r}{\bf s}} (\Delta) & = & \frac{i}{\pi\omega_0^2}
\left[ \frac{\pi}{2}\, \theta(v\Delta - |{\bf r} - {\bf s}|)
  \right. \nonumber \\ & & \left. +\ \mbox{arcsin} \left(
  \frac{v\Delta}{|{\bf r} - {\bf s}|} \right) \theta(|{\bf r} - {\bf
    s}| - v\Delta) \right].
\end{eqnarray}

Notice that the real part of the correlation function vanishes for
distances larger than $v\Delta$. For this bath as well as others, the
number of qubits within the spatial range of the correlation function
is determined by the ratio $v\Delta/a$, where $a$ is the surface code
lattice constant.

\subsection{Super-Ohmic bath}
\label{sec:superohmicbath}

Choosing $D=2$ and $s=1/2$, we find for the real part (see Appendix
\ref{sec:real_super})
\begin{equation}
\label{eq:GRsuperohmic}
{\cal G}^{(R)}_{{\bf r}{\bf s}}(\Delta) = \frac{v}{\pi \omega_{0}^{3}}
\left[ \frac{1}{|{\bf r} - {\bf s}|} - \frac{\theta(|{\bf r}-{\bf s}|
    - v\Delta)} {\sqrt{|{\bf r}-{\bf s}|^2 - (v\Delta)^2}} \right].
\end{equation}
For the imaginary part we find (see Appendix \ref{sec:imag_super})
\begin{equation}
\label{eq:GIsuperohmic}
{\cal G}^{(I)}_{{\bf r}{\bf s}} (\Delta) = \frac{iv}{\pi\omega_0^3}
\frac{\theta(v\Delta - |{\bf r}-{\bf s}|)} {\sqrt{(v\Delta)^2 - |{\bf
      r}-{\bf s}|^2}}.
\end{equation}

\begin{widetext}

\begin{figure}[t]
\includegraphics[width=.75\columnwidth]{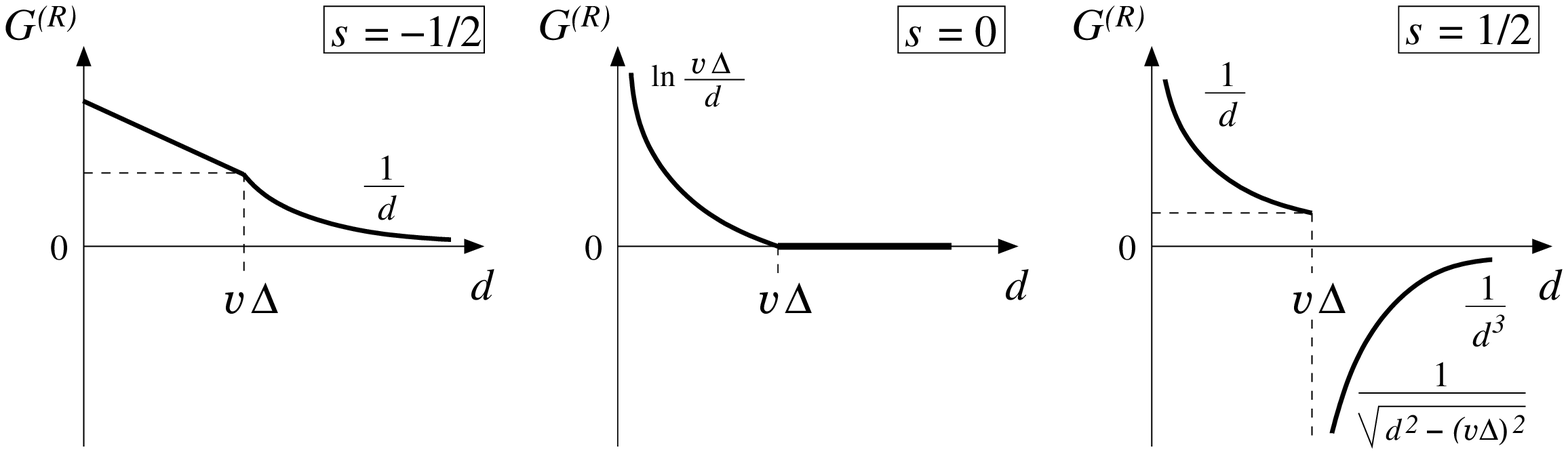}
\includegraphics[width=.75\columnwidth]{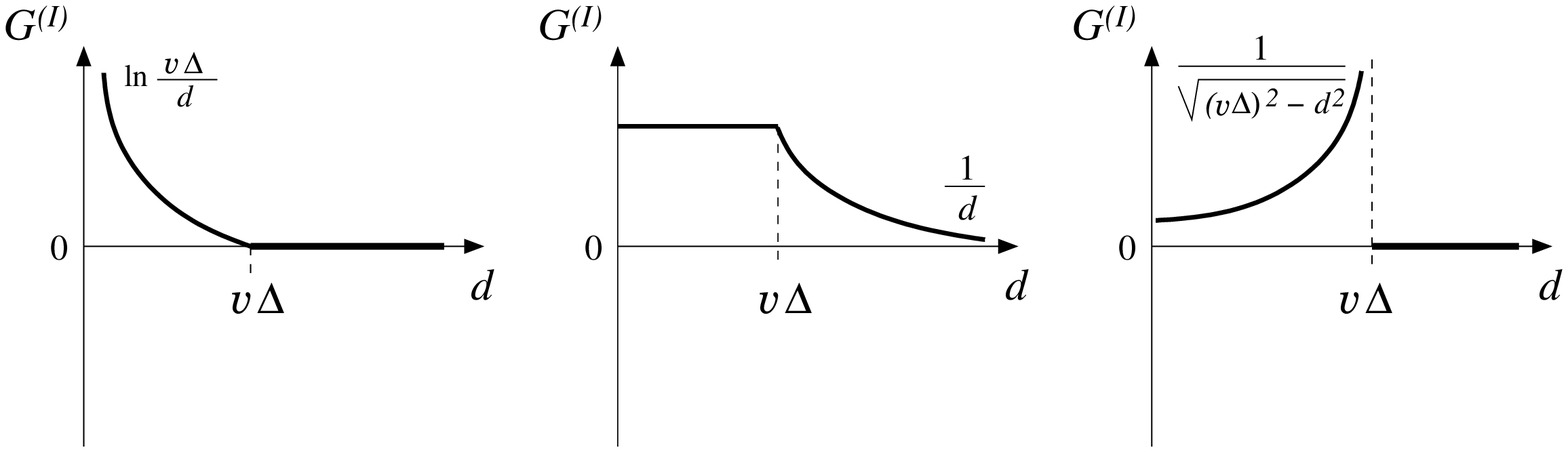}
\caption{Schematic representation of the spatial dependence of the
  correlation functions ${\cal G}^{(R)}_{{\bf r}{\bf s}}(\Delta)$ and
  ${\cal G}^{(I)}_{{\bf r}{\bf s}}(\Delta)$ for $s=-1/2$, $0$, and
  $1/2$. We use $d=|{\bf r}-{\bf s}|$.}
\label{fig:correlators}
\end{figure}

\end{widetext}

A schematic representation of the spatial dependence of these
correlations functions is shown in Fig. \ref{fig:correlators}

\section{Syndrome extraction}
\label{sec:QEC}

Let us assume that the system is prepared initially in the logical
state $\left|\bar{\uparrow}\right\rangle$ and the boson field initial
state is the vacuum,
\begin{equation}
\label{eq:Psi0}
|\Psi_{0}\rangle = \left( G|F_z \rangle \right) \otimes |0\rangle_b.
\end{equation}
We then let the system evolve under the interaction Hamiltonian until
a time $\Delta$, when an error correction protocol is performed
flawlessly. The syndrome extraction operator is equivalent to the
application of the projector
\begin{equation}
P = \frac{1}{2^{N_\square}2^{N_\lozenge}} \prod_{\square} \left( 1 +
b_{\square} B_{\square} \right) \prod_{\lozenge} \left( 1 +
a_{\lozenge} A_{\lozenge} \right),
\end{equation}
where $a_{\square}=\pm1$ and $b_{\lozenge}=\pm1$ are the syndromes for
each star and plaquette operator, respectively. Since we are assuming
only bit-flip errors, the projection over stars is just the identity
for $a_{\lozenge}=1$, namely,
\begin{equation}
P = \frac{1}{2^{N_{\square}}} \prod_{\square} \left( 1 + b_{\square}
B_{\square} \right).
\end{equation}
Using that $A_{\lozenge}G=G$ and $\left[\bar{X},G\right]=0$, we can
rewrite the projector in a slightly different form,
\begin{equation}
P = R\left| \bar{\uparrow} \right\rangle \left\langle \bar{\uparrow}
\right|R + R\left|\bar{\downarrow} \right\rangle \left\langle
\bar{\downarrow} \right|R,
\end{equation}
where $R$ is the recovery operation chosen to be performed.

In principle, we should consider all possible syndrome outcomes.
However, it is useful to look at the most benign evolution and assume
$b_{\square}=1$. Such an evolution corresponds to the system remaining
in the vacuum of the gauge fields after a time $\Delta$. This nonerror
syndrome provides an upper bound to the available computational
time. In addition, it simplifies the calculations by removing from
consideration a recovery operation that tries to steer the system back
to the computational basis \cite{poulin10,Fowler12,footnote2}. Thus,
for this particular case,
\begin{equation}
P = \left| \bar{\uparrow} \right\rangle \left\langle \bar{\uparrow}
\right| + \left|\bar{\downarrow} \right\rangle \left\langle
\bar{\downarrow} \right|.
\end{equation}

The environment state is unaffected by the error correction protocol.
If no extra step is taken to dissipate excitations that pile up over
time, the environment will keep a memory of events that happened
during the QEC period. Keeping track of such excitations between QEC
cycles in a fidelity calculation is a difficult task even for simple,
non-topological logical qubit systems \cite{novais2008, novais2010,
  Preskill2009}. For topological qubits, the task is considerably
harder due to the exponential number of terms that enter in the
composition of the computational states.

Thus, in order to make the formulation amenable to an analytical
calculation, we consider an extra step to the QEC protocol. In
addition to projecting the quantum computer wave function back to the
logical Hilbert space, we assume that at the end of the QEC step the
environment is reset to its ground state. This is equivalent to
imposing at the end of the QEC step the projector $\lim_{T_{\rm bath}
  \to 0} e^{-H_0/k_B T_{\rm bath}}$, for some environment temperature
$T_{\rm bath}$ defined with respect to some even larger reservoir.

A consequence of this extra QEC hypothesis is that we exclude from the
calculation any spatial correlation between QEC periods, as well as
memory and spatial correlations between the time evolution of bras and
kets. This new projector operator can be conveniently written as
\begin{equation}
P^{\prime} = |\Psi_{0} \rangle \langle \Psi_{0}| + \bar{X} |\Psi_{0}
\rangle \langle \Psi_{0}| \bar{X}.
\label{eq:Pprime}
\end{equation}
After the projection, the wave function must be normalized again. For
this purpose, consider the normalization factor
\begin{equation}
\left\langle \Psi_{0} \left| U^{\dagger} (\Delta) P^{\prime} U
(\Delta) \right| \Psi_{0} \right\rangle = \left| {\cal A} \right|^{2}
+ \left| {\cal B} \right|^{2},
\end{equation}
where
\begin{equation}
{\cal A} = \left\langle \Psi_{0} \left| U (\Delta) \right| \Psi_{0}
\right\rangle
\label{eq:p1}
\end{equation}
and
\begin{equation}
{\cal B} = \left\langle \Psi_{0} \left| \bar{X} U (\Delta) \right|
\Psi_{0} \right\rangle.
\label{eq:p2}
\end{equation}
Below, we use the expectation values ${\cal A}$ and ${\cal B}$ to
compute the surface code fidelity after one QEC cycle.

\section{Fidelity}
\label{sec:fidelity}

The fidelity of the surface code after one QEC cycle can be defined as
\begin{equation}
F \equiv | \langle \Psi_{{\rm QEC}} | \Psi_{0} \rangle |,
\end{equation}
where $|\Psi_0\rangle$ is the initial state of the qubit system and
the bath and
\begin{equation}
|\Psi_{{\rm QEC}} \rangle = P^{\prime} U (\Delta) \left| \Psi_{0}
\right\rangle.
\end{equation}
The expectation values ${\cal A}$ and ${\cal B}$ now come in handy
since they allow one to obtain a simple expression for the fidelity,
\begin{eqnarray}
\label{eq:fidelity}
F & = & \left[ \frac{ \left\langle\Psi_{0} \left| P^{\prime} U
    (\Delta) \right| \Psi_{0} \right\rangle \left\langle \Psi_{0}
    \left| U^{\dagger} (\Delta) P^{\prime} \right| \Psi_{0}
    \right\rangle} { \left\langle \Psi_{0} \left| U^{\dagger} (\Delta)
    P^{\prime} U (\Delta) \right| \Psi_{0} \right\rangle}
  \right]^{1/2} \nonumber \\ & = & \left[ \frac{ \left\langle \Psi_{0}
    \left| U (\Delta) \right| \Psi_{0} \right\rangle \left\langle
    \Psi_{0} \left| U^{\dagger} (\Delta) \right| \Psi_{0}
    \right\rangle} {\left| {\cal A} \right|^{2} + \left| {\cal
      B}\right|^{2}} \right]^{1/2} \nonumber \\ & = & \frac{ \left |
  {\cal A} \right|}{\sqrt{\left| {\cal A} \right|^{2} + \left|{\cal B}
    \right|^{2}}} \nonumber \\ & = & \frac{1}{\sqrt{1 + \frac{|{\cal
        B}|^2}{|{\cal A}|^2}}},
\end{eqnarray}
where we used that $P^{\prime} \left|\Psi_{0}\right\rangle =
\left|\Psi_{0}\right\rangle$. Thus, our task of determining the
fidelity is reduced to evaluating the ratio $|{\cal B}|^2/|{\cal
  A}|^2$.

\section{Mapping onto a statistical model}
\label{sec:statmodel}

Let us first consider the cases where the bath correlation function
$\Phi_{{\bf r}{\bf s}}(\Delta)$ has both real and imaginary parts
finite, namely, $0 < D+2s < 3$. The insertion of the evolution
operator given by Eq. (\ref{eq:evolution}) into Eqs. (\ref{eq:p1}) and
(\ref{eq:p2}) results in very compact expressions for ${\cal A}$ and
${\cal B}$ since these expectation values are taken on the bosonic
vacuum. After a short manipulation, we arrive at
\begin{equation}
{\cal A} = \chi \left\langle F_z \left| e^{-\beta {\cal H}}\, G^{2}
\right| F_z \right \rangle
\label{eq:A1-GPhi}
\end{equation}
and
\begin{equation}
{\cal B} = \chi \left\langle F_z \left|\bar{X} e^{-\beta {\cal H}}\,
G^{2}\right| F_z \right\rangle,
\label{eq:A2-GPhi}
\end{equation}
where we introduced
\begin{equation}
\label{eq:Heffective}
{\cal H} = \frac{\lambda^2}{2\beta} \sum_{i\neq j} \Phi_{{\bf r}_i
  {\bf r}_j} (\Delta)\, \sigma_{i}^{x} \sigma_{j}^{x}
\end{equation}
and
\begin{equation}
\label{eq:beta}
\beta = \frac{1}{2\pi} \left( \frac{\lambda}{\omega_0} \right)^2
\frac{1}{(\omega_0\Delta)^{D+2s-2}}.
\end{equation}
Clearly, Eq. (\ref{eq:Heffective}) can be interpreted as an effective
interaction between qubits intermediated by the environmental
bosons. The connection to a statistical model becomes more apparent
when we consider that the ferromagnetic state along the $z$ direction
can be expanded in the $x$ basis, namely,
\begin{equation}
\label{eq:Fz}
|F_z\rangle = \prod_{i=1}^N \left( \frac{|\uparrow\rangle_{i,x} +
  |\downarrow\rangle_{i,x}} {\sqrt{2}} \right).
\end{equation}
Inserting this expression into Eqs. (\ref{eq:A1-GPhi}) and
(\ref{eq:A2-GPhi}), we arrive at
\begin{equation}
\label{eq:AS}
{\cal A} = \frac{\chi}{2^N} \sum_S e^{-\beta E_S} \left\langle S
\left| G^2 \right| S \right\rangle
\end{equation}
and
\begin{equation}
\label{eq:BS}
{\cal B} = \frac{\chi}{2^N} \sum_S e^{-\beta E_S} \left\langle S
\left| \bar{X}\, G^2 \right| S \right\rangle,
\end{equation}
where $S$ stands for the eigenstates of the operator $\prod_{i=1}^N
\sigma_i^x$ and 
\begin{equation}
\label{eq:E_S}
\beta E_S = \langle S |\beta {\cal H} |
S\rangle. 
\end{equation}
Notice that the expectation values in Eqs. (\ref{eq:AS}) and
(\ref{eq:BS}) vanish for those states $|S\rangle$ where at least one
start operator has a $-1$ eigenvalue. Therefore, those equations can
be rewritten as
\begin{equation}
\label{eq:ASp}
{\cal A} = \frac{\chi}{2^N} \sum_{S^\prime} e^{-\beta E_{S^\prime}}
\end{equation}
and
\begin{equation}
\label{eq:BSp}
{\cal B} = \frac{\chi}{2^N} \sum_{S^\prime} e^{-\beta E_{S^\prime}}
\left\langle S^\prime \left| \bar{X} \right| S^\prime \right\rangle,
\end{equation}
where the sums are over the subset of states $S^\prime$ where {\it
  all} star operators take positive values: 
\begin{equation}
\label{eq:constraint}
\langle S^\prime | A_\lozenge | S^\prime \rangle = +1.
\end{equation}
It is clear now that ${\cal A}$ is proportional to the partition
function of a classical statistical spin model with a restricted
configuration space. Then, ${\cal C}$ is equal to the expectation
value of the operator $\bar{X}$ in this model.

The statistical model defined by Eqs. (\ref{eq:Heffective}) and
(\ref{eq:E_S}) -- (\ref{eq:BSp}) is nontrivial in a number of
ways. First, the interaction term (\ref{eq:Heffective}) is not purely
real. Second, the interaction range is not necessarily restricted to
nearest neighbors. Third, the constraint imposed by
Eq. (\ref{eq:constraint}) severely reduces the size of the
configuration space.

Since $\left\langle S^\prime \left| \bar{X} \right| S^\prime
\right\rangle$ can only take the values $\pm 1$, one can rewrite
Eqs. (\ref{eq:ASp}) and (\ref{eq:BSp}) as sums over ``energy''
eigenvalues, namely,
\begin{equation}
\label{eq:AE}
\mathcal{A} = \frac{\chi}{2^{N}} \sum_{E^\prime} \left[ g^+(E^\prime)
  + g^-(E^\prime) \right]\, e^{-\beta E^\prime}
\end{equation}
and
\begin{equation}
\label{eq:BE}
\mathcal{B} = \frac{\chi}{2^{N}} \sum_{E^\prime} \left[ g^+(E^\prime)
  - g^-(E^\prime)\right] \, e^{-\beta E^\prime},
\end{equation}
where $g^\pm(E^\prime)$ are the number of qubit configurations with
energy $E^\prime$ and $\langle\bar{X}\rangle = \pm 1$. The prime
indicates that only configurations where all star operators have $+1$
expectation value are considered, Eq. (\ref{eq:ASp}).

When the sums in Eqs. (\ref{eq:AE}) and (\ref{eq:BE}) are not
restricted by Eq. (\ref{eq:ASp}), the time-reversal symmetry of the
Hamiltonian implies $g^+(E) = g^-(E)$ for logical operators $\bar{X}$
containing an odd number of $\sigma_i^x$ qubit operators. Therefore,
in this case, $\mathcal{B}=0$. For $\bar{X}$ containing an even number
of $\sigma_i^x$ operators, for each ``energy'' eigenvalue $E$,
$\langle \bar{X} \rangle$ is either $+1$ [and $g_-(E)=0$] or $-1$ [and
  $g_+(E)=0$], but the value of ${\cal B}$ cannot be predicted.

In the case of a restricted sum, it is straightforward to see that the
separation of configurations in time-reversal symmetry classes is not
useful. Consider that at the vertical boundaries one can form
operators $A_{\lozenge}$ with three qubits. In this case, even if a
certain configuration $|S^\prime\rangle$ satisfies Eq. (\ref{eq:ASp}),
its time-reversal partner will not and therefore will not be included
in Eqs. (\ref{eq:AE}) and (\ref{eq:BE}). Thus, the restriction is
equivalent to projecting out time-reversal partner of
$|S^\prime\rangle$.

As explained in Ref. \cite{novais2013}, one useful way to understand
this point is to break up the states $\{|S^\prime\rangle\}$ into two
groups, $\{|S^\prime_+\rangle\}$ and $\{|S^\prime_-\rangle\}$, where
\begin{equation}
|S^\prime_+\rangle = \prod_j B_{\square} |F_x\rangle
\label{eq:group1}
\end{equation}
and
\begin{equation}
|S^\prime_-\rangle = \bar{Z}_\Gamma | S^\prime_+\rangle.
\label{eq:group2}
\end{equation}
Here, $\prod_j B_{\square}$ is a product of all plaquette operators
that do not touch the logical error $\bar{Z}_\Gamma$ for a given path
$\Gamma$. It is then possible to show that this separation leads to
the appearance of an effective local magnetic field that acts only on
the qubits along the path $\Gamma$. This local magnetic field leads to
the time-reversal symmetry breaking in the computation of the
expectation values in Eqs. (\ref{eq:AE}) and (\ref{eq:BE}).

\subsection{Effective interaction and fictitious temperature}
\label{sec:effective}

The parameter $\beta$ plays the role of inverse temperature in the
statistical model. From Eq. (\ref{eq:beta}), we see that $\beta$ is
proportional to $\lambda^2$, thus serving as a measure of the strength
of the coupling between the qubits and the environment. The effective
exchange interaction amplitude $J_{ij}$ depends on the bath
characteristics (e.g., spatial dimension and spectral density), on the
QEC cycle duration $\Delta$, and on the ratio $a/(\Delta v)$.

For instance, consider the Ohmic bath, where
\begin{equation}
\beta = \frac{1}{2\pi} \left( \frac{\lambda}{\omega_0} \right)^2
\end{equation}
Using Eqs. (\ref{eq:GRohmic}) and (\ref{eq:GIohmic}), the effective
Hamiltonian of the statistical model can be written as
\begin{equation}
\label{eq:HIsing}
{\cal H} = \sum_{i\neq j} J_{ij}\, \sigma_i^x \sigma_j^x,
\end{equation}
with
\begin{equation}
\label{eq:Johmic}
J_{ij} = \frac{1}{2} \times \left\{ \begin{array}{cc} \mbox{arcosh}
  \left( \frac{v\Delta}{|{\bf r}_i-{\bf r}_j|} \right) +
  \frac{i\pi}{2}, & \frac{|{\bf r}_i - {\bf r}_j|} {v\Delta} < 1,
  \\ i\, \mbox{arcsin} \left( \frac{v\Delta} {|{\bf r}_i - {\bf r}_j|}
  \right), & \frac{|{\bf r}_i - {\bf r}_j|} {v\Delta} > 1. \end{array}
\right.
\end{equation}

The real part of $J_{ij}$ is nonzero only between qubits within the
light cone of the bosonic modes. The imaginary part is present for any
pairs of qubits, but decays rapidly (approximately with the cube of
the inverse distance) when qubits are outside the light cone. For a
lattice of size $L$, an extreme limit occurs when $v\Delta\sim L$, in
which case all qubits are correlated. In the opposite limit, when the
QEC cycle period $\Delta$ is sufficiently short (or, equivalently,
that the lattice constant is large enough), so that $a/\sqrt{2} <
v\Delta < a$, only qubits belonging to the same plaquette are within
the light cone defined by the free spatial propagation of the bosonic
modes and the real part of the coefficient $J_{ij}$ vanishes beyond
nearest neighbors.

\section{Connecting $\beta$ to the qubit error probability}
\label{sec:errorprob}

It is useful to relate the fictitious inverse temperature $\beta$ of
the spin model to the probability $p$ of a qubit flipping its spin
state during the QEC cycle. The latter can be defined as
\begin{equation}
\label{eq:prob}
p = \langle 0 | \otimes \langle \uparrow_j | U^\dagger_j(\Delta)
|\downarrow_j \rangle \langle \downarrow_j | U_j(\Delta) |\uparrow_j
\rangle \otimes |0\rangle,
\end{equation}
where $\{|\uparrow_j\rangle,|\downarrow_j\rangle\}$ are states of the
qubit located at ${\bf r}_j$, $|0\rangle$ is the bath ground state,
and
\begin{equation}
\label{eq:Uj}
U_j(\Delta) = T_t \exp \left[ -\frac{\lambda}{2} \int_0^\Delta dt\,
  f({\bf r}_j,t)\, \sigma_j^x \right]
\end{equation}
is the evolution operator of that qubit coupled to the bath when the
dynamics of all other qubits is frozen. The steps in the evolution of
Eq. (\ref{eq:prob}) are similar to those used in the derivation of the
fidelity. The details are provided in Appendix
\ref{sec:AppendixD}. The result is
\begin{equation}
\label{eq:probexp}
p = \frac{1}{2} \left\{ 1 - \exp \left[ - \frac{\lambda^2}{4}\, {\cal
    G}^{(R)}_{{\bf r}_j{\bf r}_j}(\Delta) \right] \right\}.
\end{equation}
Notice that for $\lambda=0$, $p=0$. As the coupling between qubits and
bath grows in magnitude, $p$ approaches 1/2, which signals a complete
randomization of the qubit state.

The functional relation between $p$ and the fictitious temperature
$\beta$ can be easily established by evoking Eq. (\ref{eq:beta}) and
employing the explicit form of ${\cal G}^{(R)}_{{\bf r}_j{\bf
    r}_j}(\Delta)$ as given in Eq. (\ref{eq:GRexplicit}). One obtains
\begin{equation}
\ln \left( 1 - 2p \right) = -\frac{\pi\beta (v\Delta)^{D+2s-2}}{L^D}
\sum_{{\bf k}\neq 0} |{\bf k}|^{2s-2} [1-\cos(|{\bf k}|v\Delta)].
\end{equation}
The sum over momentum diverges in the ultraviolet when $D+2s\geq 2$
and the relation between the error probability and the fictitious
inverse temperature becomes cutoff dependent. For instance, for $D=2$
and $s=0$ (Ohmic bath), one finds $p = \frac{1}{2}\left[ 1-
  (2v\Delta\Lambda)^{-\beta/2} \right]$, where $\Lambda$ is the
ultraviolet momentum cutoff. However, for $D=2$ and $s=-1/2$
(sub-Ohmic case), one finds $p = \frac{1}{2}\left(1-e^{-\pi \beta/4}
\right)$, which is cutoff independent.

\section{Estimate of $\beta_c$}
\label{sec:estimate}

We now provide an estimate of the critical inverse fictitious
temperature, taking a slightly different approach from that used in
Ref. \cite{novais2013}. Let us consider the case when the effective
spin coupling in Eq. (\ref{eq:Heffective}) is real and only
nearest-neighbor interactions occur, $\Phi_{{\bf r}_i{\bf
    r}_j}(\Delta)=J$ for $|{\bf r}_i-{\bf r}_j| \leq a/\sqrt{2}$ and
$\Phi_{{\bf r}_i{\bf r}_j}(\Delta)=0$ otherwise. We will carry out a
low-temperature (large $\beta$) expansion of Eqs. (\ref{eq:AE}) and
(\ref{eq:BE}). The key element we exploit in this expansion is the
following property of the surface code: At the boundaries of the
surface code, the star operators are defined by three qubits instead
of four. This means that if a certain state $|S \rangle$ is included
in the restricted sums defining ${\cal A}$ and ${\cal B}$, its
time-reversed counterpart is included. This is because reversing all
the qubit of a configuration with an eigenvalue $+1$ for all star
operators yields a $-1$ eigenvalue for the star operators at the
boundaries.

This property is particularly useful in the limit of $\beta
\rightarrow \infty $ when the term with the minimum energy, $e^{-\beta
  E^\prime_{\rm min}}$, carries the leading contribution to the
sums. For $\beta \rightarrow \infty $ and for $J<0$, the only state
with minimum energy is a ferromagnetic state where all spins are
pointing along the positive $x$ direction, $|F_x\rangle$. Since the
the ferromagnet state with spins pointing along the negative $x$ is
not part of the restricted sum, $g^-(E^\prime_{\rm min}) = 0$. In this
limit, the spin model is in the ordered phase, with ${\cal A} = {\cal
  B}$, resulting in ${\cal F} = 1/2$ (lost fidelity).

Consider now a large but finite $\beta$. Starting from the state $|F_x
\rangle$, the states appearing in $\cal A$ and $\cal B$ can be
separated into two groups, as shown in Eqs. (\ref{eq:group1}) and
(\ref{eq:group2}). The first group, $\{S_+^\prime\}$, corresponds to
the states counted in $g^+(E^\prime)$, whereas the second group,
$\{S_+^\prime\}$, is accounted for by $g^-(E^\prime)$. Therefore,
$g^-(E^\prime)$ is equal to the number of states of energy $E^\prime$
with a logical error $\bar{Z}_\Gamma$ for a given path $\Gamma$. At
large $\beta$, the energy cost of these states is of the order of the
length of $\bar{Z}_\Gamma$ and they are suppressed in comparison to
other states. The leading terms contributing to the sums are the
minimum energy state $|F_x\rangle$ and the states $|S^\prime_+\rangle$
containing only small loops. However, as the system size increases the
multiplicity factor $g^+( E^\prime)$ increases as well. Its value is
proportional to the number of ways one can apply the $\bar{Z}$
operation, or equivalently, to the number of self-avoiding walks (SAW)
from one open boundary to its opposite. The number of SAWs with a
length $l$, is related to connective constant $\mu$ of the lattice and
scales as $\mu^{l}$.

If the multiplicity factor $ g^-(E^\prime)$ is high enough, it can
compensate the Boltzmann factor suppression. Then, as $\beta$
decreases to a certain value $\beta_c$, for some energy
$E^\prime_\star$ a term of the type $\mu^{l_{\bar{Z}}(E^\prime_\star)}
\, e^{-\beta_ {c}E_\star^\prime}$ will appear in ${\cal B}$ with the
same order as the leading term related to the $|S^\prime_+\rangle$
states, namely, $e^{-\beta_ {c}E^\prime_{\rm min}}$. Here,
$l_{\bar{Z}}$ is the length of $\bar{Z}$. This criterion provides a
crude estimate for $\beta_c$:
\begin{equation}
\label{eq:Estimation}
e^{-\beta_c E^\prime_{\rm min}} \approx
\mu^{l_{\bar{Z}}(E^\prime_\star)} \, e^{-\beta_c E^\prime_\star},
\end{equation}
leading to
\begin{equation}
\label{eq:beta_cestim}
\beta_c \approx \frac{l_{\bar{Z}}(E^\prime_\star)\, \ln
  \mu}{E^\prime_{\star}- E^\prime_{\rm min}}.
\end{equation}
The difference between $E^\prime_{\star}$ and $E^\prime_{\rm min}$ is
proportional to the length of the logical error $\bar{Z}$. Then, the
denominator is of the order of $2n\, l_{\bar{Z}} J$, where $n$ is the
number of qubits interacting with the qubits comprising the logical
error $\bar{Z}$ through the Hamiltonian in Eq. (\ref{eq:HIsing}). A
range of possible values for the connective constant of a square
lattice can be found in the literature. If we adopt $\mu=2.64$
\cite{madras}, set $n=4$, and insert these values into
Eq. (\ref{eq:beta_cestim}), we obtain $\beta_c J \approx 0.12$.

\section{Numerical evaluation of the fidelity}
\label{sec:numerics}

In light of Sec. \ref{sec:statmodel}, we can interpret the ratio
${\cal B}/{\cal A}$ as the expectation value of the $M_x = \sum_{j\in
  \Omega} \sigma_j^x$, namely, the $x$ magnetization of a linear set
of spin 1/2 particles embedded into a spin system governed by the
Hamiltonian of Eq. (\ref{eq:HIsing}) with the restriction imposed by
Eq. (\ref{eq:ASp}). (Here, $\Omega$ denotes the path defining the
logical error $\bar{X}$.) In the absence of such a restriction, the
computation of ${\cal B}/{\cal A}$ in the thermodynamic limit would
follow standard procedures used in statistical mechanics. The
restriction, however, makes an analytical computation rather
difficult, if not impossible. Therefore, we resort to numerical
calculations, both exact and approximate, to find how ${\cal B}$ and
${\cal A}$ (and thus the fidelity ${\cal F}$) behave as a function of
$\beta$ and how this behavior scales with increasing system sizes.

Below, we focus on the case where the effective interaction strength
$J_{ij}$ is real and only involves nearest-neighbor qubits. As
mentioned earlier, this special case is of significance to experiments
where $v\Delta$ is of the order of $a$. Short-range correlated errors
in this case can be introduced by any measurement or operation on
individual stars and plaquettes.

\subsection{Exact calculations}
\label{sec:exactnumerics}

\begin{figure}[t]
\includegraphics[width=0.9\columnwidth]{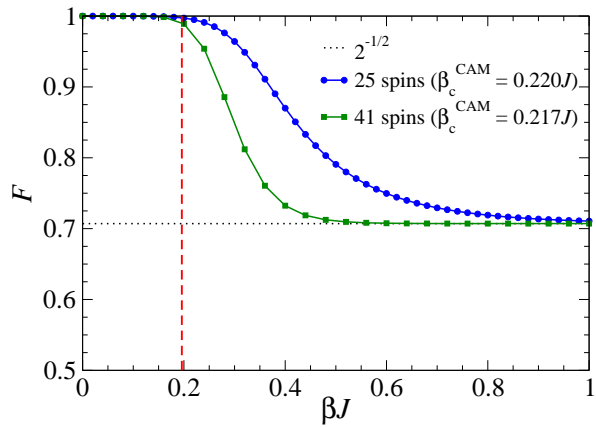}
\caption{(Color online) Surface code fidelity of code spaces of 25 and
  41 physical qubits in contact with a bosonic bath when star
  operators are restricted to positive values ($A_{\lozenge}=1$).}
\label{fig:Fidelity}
\end{figure}

For two lattice sizes, $N=25$ and 41, we computed $\mathcal{A}$ and
$\mathcal{B}$ for an $\bar{X}$ operator that ran vertically through
the middle of the lattice. The computation was done by exhaustive
enumeration of all orthogonal qubit configurations $|S\rangle$ that
complied with the constraint $\langle S|G^2|S\rangle \neq 0$, namely,
that produced only positive plaquette eigenvalues. We verified that
the results were insensitive to the choice of operator $\bar{X}$. The
resulting fidelity is shown in Fig. \ref{fig:Fidelity} as a function
of the inverse fictitious temperature $\beta$. For small $\beta$
(equivalent to small coupling constant $\lambda$), the fidelity stays
close to 1 after one QEC cycle. As $\beta$ increases, the fidelity
decays and tends asymptotically to $1/\sqrt{2}$, which is expected when
${\cal B}=0$. Another important feature is that the transition from
${\cal F}=1$ to ${\cal F}=1/\sqrt{2}$ becomes sharper as the system
size is increased. This is the expected behavior when, in the
thermodynamic, infinite-size limit, a phase transition occurs at some
critical value of $\beta$.

\begin{figure}[t]
\includegraphics[width=0.9\columnwidth,scale=0.4]{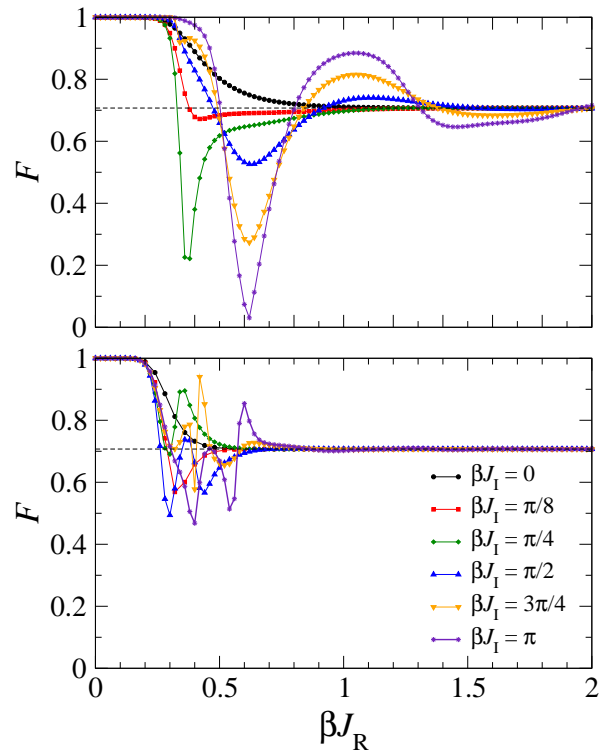}
\caption{(Color online) Fidelity of a code space of 25 physical qubits
  in contact with a bosonic bath when star operators are restricted to
  positive values ($A_{\lozenge}=1$) and an imaginary part is added to
  the coupling constant: $J=J_R+iJ_I$. The data sets correspond to
  different values of $J_I$.}
\label{fig:Fidelity_complex}
\end{figure}

We have tested that this behavior is not substantially altered when
the coupling constant $J$ gains a constant imaginary part. The results
are shown in Fig. \ref{fig:Fidelity_complex}. The main effect of
adding an imaginary part is to create oscillations in the decay of the
fidelity as a function of $\beta$. The larger the magnitude of the
imaginary part in $J$, the more oscillations are observed. However,
the relative amplitude of these oscillations decrease with increasing
system size. In the limit of a large number of physical qubits, we
expect the oscillations to be relatively small and concentrated near
the critical value $\beta_c$.

In order to determine the critical value $\beta_c$, we resort to the
coherent anomaly method, which has been extensively and successfully
used to determine critical temperatures in interacting spin systems
\cite{suzuki86,suzuki87}.

\subsection{Mean-field solution: Coherent anomaly method}
\label{sec:CAM}

In the coherent anomaly method (CAM), a cluster of interacting spins
is embedded inside a mean-field medium. Self-consistency is obtained
by allowing the spins at the boundary of the cluster to experience the
mean field, which is set equal to the mean value of the central spin
in the cluster. This constraint provides an equation from which the
critical temperature can be determined. As the cluster size is
increased, the expectation is that the critical temperature obtained
in this way rapidly approaches the exact value of an infinite-size
system \cite{suzuki86,suzuki87}.

More precisely, let $S_0^x$ denote the central spin operator and let
${\cal H}$ be a Hamiltonian describing nearest-neighbor interactions
inside the cluster, ${\cal H}_{\rm cl}$, as well as the action of an
effective field $\phi_{\rm eff}$ at the boundary spins,
\begin{equation}
\label{eq:HCAM}
{\cal H} = {\cal H}_{\rm cal} + J\, \phi_{\rm eff} \sum_{i \in
  \partial \Omega} S_i^x,
\end{equation}
where $\partial\Omega$ denotes the cluster boundary. The expectation
value of the central spin is given by
\begin{equation}
\langle S_0^x \rangle = \frac{\mbox{Tr} \left[ S_0^x\, e^{-\beta {\cal
        H}} \right]} {\mbox{Tr} \left[ e^{-\beta{\cal H}} \right]},
\end{equation}
where the trace is carried over all allowed spin
configurations. Expanding the exponentials in Eq. (\ref{eq:HCAM}) to
the lowest nontrivial order in the effective field, we find that
\begin{equation}
\langle S_0^x \rangle = \langle S_0^x \rangle_{\rm cl} - \beta\, J\,
\phi_{\rm eff} \sum_{i\in \partial\Omega} \langle S_0^x S_i^i
\rangle_{\rm cl},
\end{equation}
where $\langle \cdots \rangle_{\rm cl}$ denotes the expectation value
taken with just the Hamiltonian ${\cal H}_{\rm cl}$ and neglecting the
boundary field. Setting $\langle S_0^x \rangle$ equal to $\phi_{\rm
  eff}$, we arrive at the equation
\begin{equation}
\label{eq:selfconsistency}
1 - \beta_c\, J \sum_{i\in \partial\Omega} \langle S_0^x S_i^x
\rangle_{\rm cl} = 0,
\end{equation}
which can be solved numerically to yield the critical inverse
temperature $\beta = \beta_c$ as a function of $J$. The most costly
part of the procedure is the calculation of the correlation function
$\langle S_0^x S_i^x \rangle_{\rm cl}$, which requires an exhaustive
enumeration of all spin configurations within the cluster.

\begin{figure}[t]
\includegraphics[width=0.9\columnwidth,scale=0.4]{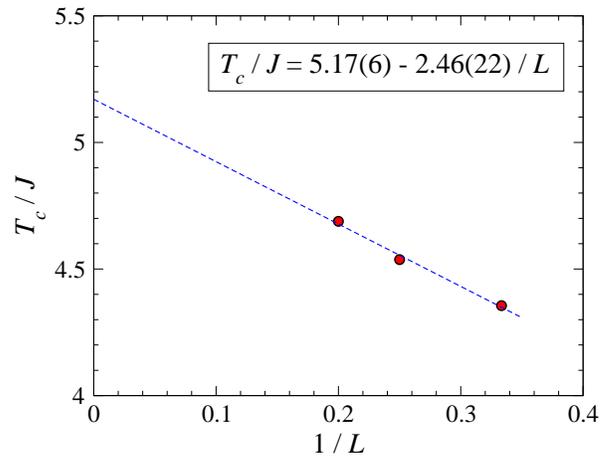}
\caption{(Color online) Finite-size scaling of the critical fictitious
  temperature $T_c$ obtained from cluster mean-field calculations for
  lattice of sizes 13, 25, and 41. A real Ising interaction of
  strength $J$ involving only nearest neighbors was used. The circles
  are the numerical data and the dashed line is a linear fit. $L$
  denotes the linear size of the surface code.}
\label{fig:extrapolation}
\end{figure}

We employed this method to compute the critical value of $\beta$ for
surface codes with clusters of increasing sizes and performed a
finite-scaling analysis to estimate the critical value in the
thermodynamic limit. The result is shown in
Fig. \ref{fig:extrapolation}. As in the case of the exact numerical
calculations, we employed the constrained nearest-neighbor Ising model
of Sec. \ref{sec:statmodel} with $J$ real and only allowed for spin
configurations with positive plaquette eigenvalues. We find that
$\beta_c J = 0.193(2)$ for an infinitely large system, which is about
60\% higher than the estimate presented in Sec. \ref{sec:estimate}
(given the roughness of the approximations involved in the estimate,
the discrepancy seems quite acceptable). The extrapolated value also
matches quite closely the point where the downturn of the fidelity
develops (see Figs. \ref{fig:Fidelity} and
\ref{fig:Fidelity_complex}), providing additional support for the
existence of a phase transition in the thermodynamic limit in the case
of nearest-neighbor interactions.

\section{Discussion and conclusions}
\label{sec:conclusions}

We have presented a fully quantum mechanical calculation of the
fidelity of the surface code in the presence of a bosonic bath. We
considered the fidelity after a complete quantum error correction
cycle and in the most benign case, when a nonerror syndrome occurs. An
important assumption made in the calculation was the resetting of the
bath to its ground state after the syndrome extraction. We then
expressed the fidelity as a function of the ratio of two complex
amplitudes which were formulated as expectation values of a
statistical spin model with complex two-body interactions and a
restricted configuration space. We presented both analytical estimates
and numerical evidence that the statistical spin model sustains an
ordered to disordered phase transition in the limit of an infinite
number of qubits. The existence of such a phase transition can be
directly related to the existence of a threshold on resilience of the
surface code: provided that the bath coupling constant remains lower
than a critical value, fidelity can be maintained close to unity with
increasing system size. This is very good news for those interested in
large-scale implementations of the surface code.

This work provides more detailed derivations and deeper analyses than
Ref. \cite{novais2013}, in additional to numerical support to the
existence of an error threshold in the surface code in the presence of
correlated noise. We note that our approach differs substantially from
the surface code literature since we do not rely on a stochastic error
model.

In order to understand the difference between a threshold due to
correlations and the usual stochastic model discussion, let us
consider the case of $D=2$ and $s=0$ (Ohmic bath). As shown in
Sec. \ref{sec:errorprob}, the probability of such a bosonic bath to
produce a bit flip error is $p = \frac{1}{2}\left[ 1-
  (2v\Delta\Lambda)^{-\beta/2} \right]$. If we take the ultraviolet
cutoff to infinity, then we are bound to find $p=1/2$. However, in
most physics systems, and condensed matter systems in particular, the
cutoff is finite. Hence, we can expand $p$ for small $\beta$ and
$\Delta$ to obtain
\begin{equation}
\beta \approx \frac{4p}{\ln\left|2v\Delta\Lambda\right|}.
\end{equation}
For the Ohmic model, we also found that the real part of the effective
interaction is given by [see Eq. (\ref{eq:Johmic})]
\begin{equation}
J_{ij} \approx \frac{1}{2} \ln\left( \frac{v\Delta}{\left|
  \mathbf{r}_i- \mathbf{r}_j\right|} \right)
\end{equation}
for qubits within the causality cone. In both equations, the
logarithms are slow growing functions and should be regarded as
producing numbers of the same order. Therefore, we can rewrite $p$ as
\begin{equation}
\beta \approx \frac{2 p}{J},
\end{equation}
where we took $J \sim J_{ij} \sim (1/2)\ln \left| 2v\Delta\Lambda
\right|$. Now, we have also found that the inverse critical
temperature is [see Eq. (\ref{eq:beta_cestim})]
\begin{equation}
\beta_c \approx \frac{\ln \mu}{2n J}.
\end{equation}
To be resilient to correlated errors, we must require the system to be
above the critical temperature, thus $\beta < \beta_c$. Using the
equations above, we find that
\begin{equation}
p < \frac{\ln \mu}{4 n}.
\end{equation}
For the least correlated case, where only nearest-neighbors effective
interactions take place, we have
\begin{equation}
p < \frac{\ln \mu}{16} \sim 6\%.
\end{equation}
That is, the threshold for the surface code in the presence of an
Ohmic bath is reduced to at most $p_c \sim 6\%$ due to the
introduction of nearest-neighbors correlated errors. If we allow for
longer-range correlated errors, the threshold will steadily decline.

Within our formulation, exact analytical calculations of the threshold
based on the fidelity are daunting. Thus, it is likely that
quantitative results will always require numerical simulations. We are
in the process of simulating statistical spin models with more general
interactions than the nearest-neighbor case investigated here. In
addition, further investigations are necessary to relax the assumption
of bath resetting and to evaluate the effects of residual qubit
correlations on the fidelity over multiple cycles. Such studies are
also under way.

\begin{acknowledgments}

We acknowledge insightful conversations with R. Raussendorf and thank
him for the hospitality at the University of British Columbia, Canada,
where this work was initiated. E.N. and E.M. also thank L. G. Dias at
the Universidade de S\~ao Paulo, Brazil, for his hospitality. E.N. was
partially supported by INCT-IQ and CNPq (Brazil). This work was
supported in part by the Office of Naval Research and the National
Science Foundation (USA).

\end{acknowledgments}

\appendix

\section{Evolution operator}
\label{sec:appendixA}

Consider Eq. (\ref{eq:UDelta}), where the bosonic field in the
interaction picture reads
\begin{eqnarray}
\label{eq:frt}
f\left({\bf r},t\right) & = & \frac{(v/\omega_{0})^{D/2+s}}{L^{D/2}}
\sum_{{\bf k}\neq0} \left|{\bf k} \right|^{s} \left( e^{i{\bf
    k}\cdot{\bf r}-i\left|{\bf k} \right| vt} a_{{\bf k}}^{\dagger}
\right. \nonumber \\ & & \left. +\ e^{-i{\bf k}\cdot{\bf
    r}+i\left|{\bf k} \right| vt} a_{{\bf k}}\right).
\end{eqnarray}
We can write $U ( \Delta) = \exp [ \Omega (\Delta)]$, where $\Omega
(\Delta)$ follows the Magnun expansion
\begin{equation}
  \Omega ( \Delta) = \Omega_1 ( \Delta) + \Omega_2 ( \Delta) + \Omega_3 (
  \Delta) + \cdots
\end{equation}
with
\begin{equation}
\Omega_1 ( \Delta) = - \frac{i\lambda}{2} \int_0^{\Delta} d t \sum_i f
({\bf r}_i, t)\, \sigma_i^x,
\end{equation}
\begin{eqnarray}
\Omega_2 ( \Delta) & = & -\frac{1}{2!} \left( \frac{\lambda}{2}
\right)^2 \int_0^{\Delta} d t_1 \int_0^{t_1} d t_2 \nonumber \\ & &
\times \sum_{i, j} [ f ({\bf r}_i, t_1), f ({\bf r}_j, t_2)]\,
\sigma_i^x \sigma_j^x,
\end{eqnarray}
\begin{eqnarray}
\Omega_3 ( \Delta) & = & -\frac{i}{3!}  \left( \frac{\lambda}{2}
\right)^3 \int_0^{\Delta} d t_1 \int_0^{t_1} d t_2 \int_0^{t_2} d t_3
\nonumber \\ & & \times \sum_{i, j, k} \big( [ f ({\bf r}_i, t_1), [ f
    ({\bf r}_j, t_2), f ({\bf r}_k, t_3)]] \nonumber \\ & & +\ [ f
  ({\bf r}_k, t_3), [ f ({\bf r}_j, t_2), f ({\bf r}_i, t_1)]]\, \big)
\nonumber \\ & & \times\, \sigma_i^x \sigma_j^x \sigma_k^x,
\end{eqnarray}
etc. Since $[f ({\bf r}_i, t_1), f ({\bf r}_j, t_2)]$ is a $c$ number,
only the first two terms in the expansion survive and we can write
\begin{eqnarray}
  U (\Delta) & = & \exp \left[ - i \frac{\lambda}{2} \sum_i F_{{\bf
        r}_i} (\Delta) \sigma_i^x \right] \nonumber \\ & &
  \times\,\exp \left[ - \frac{\lambda^2}{8} \sum_{i, j}
    \mathcal{G}^{(I)}_{{\bf r}_i {\bf r}_j} (\Delta) \sigma_i^x
    \sigma_j^x \right].
\label{eq:13}
\end{eqnarray}
where $\mathcal{G}^{(I)}_{{\bf r}_i {\bf r}_j} (\Delta)$ and $F_{{\bf
    r}_i} (\Delta)$ are defined in Eqs. (\ref{eq:I}) and (\ref{eq:F}),
respectively.

It is convenient to rewrite the first exponential in Eq. (\ref{eq:13})
as a normal ordered term. For this purpose, we note that $F_{{\bf
    r}_i} (\Delta)$ has the form
\begin{equation}
  F_{{\bf r}_i} (\Delta) = \sum_{k \neq 0} ( g_{{\bf r}_i, {\bf
      k}}^{\ast}\, a_{{\bf k}} + g_{{\bf r}_i, {\bf k}}\, a_{{\bf
      k}}^{\dagger}),
\label{eq:14}
\end{equation}
where
\begin{equation}
g_{{\bf r}, {\bf k}} = - i \frac{(v/\omega_0)^{D/2+s}}{v L^{D / 2}}
|{\bf k}|^{s - 1} e^{i{\bf k} \cdot {\bf r}} ( e^{i |{\bf k}| v
  \Delta} - 1).
\label{eq:15}
\end{equation}
Thus, we can write
\begin{equation}
\exp \left[ - \frac{i\lambda}{2} \sum_i F_{{\bf r}_i} (\Delta)
  \sigma_i^x \right] = \prod_{{\bf k} \neq 0} \exp ( a_{{\bf
    k}}^{\dagger} v_{{\bf k}} - a_{{\bf k}} v_{{\bf k}}^{\dagger}),
\end{equation}
where
\begin{equation}
v_{{\bf k}} = - \frac{i\lambda}{2} \sum_i g_{{\bf r}_i, {\bf k}}\,
\sigma_i^x .
\end{equation}
Now we can use the Zassenhaus formula applied to bosonic operators,
\begin{equation}
e^{(a_{{\bf k}}^{\dagger} v_{{\bf k}} - a_{{\bf k}} v_{{\bf
      k}}^{\ast})} = e^{\frac{1}{2} [ a_{{\bf k}}^{\dagger} v_{{\bf
        k}}, a_{{\bf k}} v_{{\bf k}}^{\dagger}]}\, e^{ a_{{\bf
      k}}^{\dagger} v_{{\bf k}}}\, e^{ - a_{{\bf k}} v_{{\bf
      k}}^{\dagger}},
\end{equation}
which results in
\begin{eqnarray}
\label{eq:normalorder}
  \exp \left[ - i \lambda \sum_i F_{{\bf r}_i} (\Delta)\, \sigma_i^x
    \right] & = & \prod_{{\bf k} \neq 0} e^{ - \frac{1}{2} v_{{\bf k}}
    v_{{\bf k}}^{\dagger}}\, e^{a_{{\bf k}}^{\dagger} v_{{\bf k}}}\,
  e^{- a_{{\bf k}} v_{{\bf k}}^{\dagger}} \nonumber \\ & = &
  \prod_{{\bf k} \neq 0} e^{- \frac{1}{2} v_{{\bf k}} v_{{\bf
        k}}^{\dagger}} : e^{a_{{\bf k}}^{\dagger} v_{{\bf k}} -
    a_{{\bf k}} v_{{\bf k}}^{\dagger}} : \nonumber \\ & = &
  e^{-\frac{1}{2} \sum_{{\bf k} \neq 0} v_{{\bf k}} v_{{\bf
        k}}^{\dagger}} \nonumber \\ & & \times : \exp \left[ - i
    \frac{\lambda}{2} \sum_i F_{{\bf r}_i} (\Delta)\, \sigma_i^x
    \right] :\, ,
\end{eqnarray}
where $:(\ldots):$ denotes normal ordering. We can rewrite the
argument of the first exponential in Eq. (\ref{eq:normalorder}) since
\begin{eqnarray}
\label{eq:vv}
  \sum_{{\bf k} \neq 0} v_{{\bf k}} v_{{\bf k}}^{\dagger} & = & -
  \frac{\lambda^2}{4} \sum_{i, j} \left( \sum_{{\bf k} \neq 0} g_{{\bf
      r}_i, {\bf k}} g_{{\bf r}_j, {\bf k}}^{\ast} \right) \sigma_i^x
  \sigma_j^x \nonumber \\ & = & - \frac{\lambda^2}{4} \sum_{i, j}
        {\cal G}^{(R)}_{{\bf r}_i{\bf r}_j}(\Delta)\, \sigma_i^x
        \sigma_j^x,
\end{eqnarray}
where ${\cal G}^{(R)}_{{\bf r}_i{\bf r}_j}(\Delta)$ is defined in
Eq. (\ref{eq:GR}). Combining Eqs. (\ref{eq:13}),
(\ref{eq:normalorder}), and (\ref{eq:vv}), we arrive at
Eq. (\ref{eq:evolution}).

\section{Correlator ${\cal G}^{(R)}_{{\bf r}{\bf s}}(\Delta)$}
\label{sec:appendixB}

The correlator in Eq. (\ref{eq:GR}) can be evaluated in the following
way. Inserting Eq. (\ref{eq:F}) into Eq. (\ref{eq:GR}) and using
Eq. (\ref{eq:frt}), we find
\begin{equation}
\label{eq:GRexplicit}
{\cal G}^{(R)}_{{\bf r}{\bf s}} (\Delta) =
\frac{(v/\omega_0)^{D+2s}}{v^2 L^D} \sum_{{\bf k}\neq 0} |{\bf
  k}|^{2s-2} e^{i {\bf k} \cdot ({\bf r}-{\bf s})} \left| e^{i|{\bf
    k}| v\Delta}-1 \right|^2.
\end{equation}
This is integral is convergent provided that $0 < D + 2s <
4$. Assuming $D=2$, we can write
\begin{eqnarray}
{\cal G}^{(R)}_{{\bf r}{\bf s}} (\Delta) & = &
\frac{2(v/\omega_0)^{2+2s}}{v^2} \int \frac{d^2k}{(2\pi)^2} |{\bf
  k}|^{2s-2} e^{i {\bf k} \cdot ({\bf r}-{\bf s})} \nonumber \\ & &
\times \left[ 1 - \cos \left( |{\bf k}| v \Delta \right) \right]
\nonumber \\ & = & \frac{(v/\omega_0)^{2+2s}}{\pi\, v^2}
\int_0^\Lambda dk\, k^{2s-1} J_0 \left( k|{\bf r} - {\bf s}|\right)
\nonumber \\ & & \times \left[ 1 - \cos \left( k v \Delta \right)
  \right],
\end{eqnarray}
where $J_n(x)$ is the $n$th Bessel function of the first kind. To
proceed further, we consider three representative values of $s$ where
the integration over momentum is convergent independent of the cutoff
and we can set $\Lambda\rightarrow\infty$.

\subsection{Sub-Ohmic case}
\label{sec:real_sub}

For $s=-1/2$ we can write
\begin{equation}
{\cal G}^{(R)}_{{\bf r}{\bf s}} (\Delta) = \frac{1}{\pi\, v\,
  \omega_0} \int_0^\infty \frac{dk}{k^2} J_0 \left( k|{\bf r} - {\bf
  s}|\right) \left[ 1 - \cos \left( k v \Delta \right) \right].
\end{equation}
Then, Eq. (\ref{eq:GRsubohmic}) can be obtained by using the integral
\cite{gradshteyn_Rsubohmic}
\begin{eqnarray}
& & \int_0^\infty \frac{dx}{x^2} J_0(\beta x)\, [1-\cos(\alpha x)]
  \nonumber \\ & = & -\beta + \left\{ \begin{array}{ll} \sqrt{\beta^2
      - \alpha^2} + \alpha\, \mbox{arcsin} \left( \frac{\alpha}{\beta}
    \right), & \alpha < \beta, \\ \frac{\alpha\, \pi}{2}, & \alpha >
    \beta. \end{array} \right.
\end{eqnarray}
%

\subsection{Ohmic case}
\label{sec:real_ohmic}

For $s=0$ we can write
\begin{equation}
{\cal G}^{(R)}_{{\bf r}{\bf s}} (\Delta) = \frac{1}{\pi\, \omega_0^2}
\int_0^\infty \frac{dk}{k}\, J_0 \left( k|{\bf r} - {\bf s}|\right)
\left[ 1 - \cos \left( k v \Delta \right) \right].
\end{equation}
Then, Eq. (\ref{eq:GRohmic}) can be obtained by using the integral
\cite{gradshteyn}
\begin{equation}
\int_0^\infty \frac{dx}{x} J_0(\beta x)\, [1-\cos(\alpha x)] =
\mbox{arcosh} \left( \frac{\alpha}{\beta} \right)\,
\theta(\alpha-\beta).
\end{equation}
%

\subsection{Super-Ohmic case}
\label{sec:real_super}

For $s=1/2$ we can write
\begin{equation}
{\cal G}^{(R)}_{{\bf r}{\bf s}} (\Delta) = \frac{v}{\pi\, \omega_0^3}
\int_0^\infty dk\, J_0 \left( k|{\bf r} - {\bf s}|\right) \left[ 1 -
  \cos \left( k v \Delta \right) \right].
\end{equation}
Then, Eq. (\ref{eq:GRsuperohmic}) can be obtained by using the
integral \cite{gradshteyn_Rsuperohmic}
\begin{equation}
\int_0^\infty dx\, J_0(\beta x)\, \left[ 1 - \cos(\alpha x) \right]=
\frac{1}{\beta} - \frac{\theta(\beta-\alpha)}{\sqrt{\beta^2-\alpha^2}}.
\end{equation}
%

\section{Correlator ${\cal G}^{(I)}_{{\bf r}{\bf s}}(\Delta)$}
\label{sec:appendixC}

\begin{widetext}
The correlator in Eq. (\ref{eq:I}) can be evaluated in the following
way. Starting with Eq. (\ref{eq:ftr}), we have
\begin{equation}
[ f ({\bf r}, t_1), f ({\bf s}, t_2)] = - 2 i
\frac{(v/\omega_0)^{D+2s}}{L^D} \sum_{{\bf k} \neq 0} |{\bf k}|^{2 s}
\sin [ {\bf k} \cdot ( {\bf r}-{\bf s}) + |{\bf k}| v ( t_1 - t_2)],
\end{equation}
which allows us to write
\begin{equation}
\label{eq:ff}
\frac{1}{2} \left\{ [ f ({\bf r}, t_1), f ({\bf s}, t_2)] + [ f ({\bf
    s}, t_1), f ({\bf r}, t_2)] \right\} = - 2 i
\frac{(v/\omega_0)^{D+2s}}{L^D} \sum_{{\bf k} \neq 0} |{\bf k}|^{2 s}
\cos [ {\bf k} \cdot ( {\bf r}-{\bf s})] \sin[ |{\bf k}| v ( t_1 -
  t_2)].
\end{equation}
Considering $D=2$, we have
\begin{equation}
\label{eq:ff2}
\frac{1}{2} \left\{ [ f ({\bf r}, t_1), f ({\bf s}, t_2)] + [ f ({\bf
    s}, t_1), f ({\bf r}, t_2)] \right\} = - \frac{i}{\pi} \left(
\frac{v}{\omega_0} \right)^{2+2s}\int_0^\Lambda dk\, k^{2 s+1} J_0 (k
|{\bf r}-{\bf s}|) \sin[ k v (t_1 - t_2)].
\end{equation}
In Eq. (\ref{eq:ff2}), we introduced an ultraviolet momentum cutoff
$\Lambda$. To proceed further, we need to specify $s$. Below, we
consider three representative values.
%

\subsection{Sub-Ohmic case}
\label{sec:imag_sub}

For $s=-1/2$, the integral in Eq. (\ref{eq:ff2}) converges. Using the
integral \cite{gradshteyn_Isubohmic}
\begin{equation}
\label{eq:GR6.761.7}
\int_0^\infty dx\, J_0(\beta x)\, \sin(\alpha x) =
\mbox{sgn}(\alpha)\, \frac{\theta(|\alpha| -
  |\beta|)}{\sqrt{\alpha^2-\beta^2}},
\end{equation}
we have
\begin{equation}
\frac{1}{2} \left\{ [ f ({\bf r}, t_1), f ({\bf s}, t_2)] + [ f ({\bf
    s}, t_1), f ({\bf r}, t_2)] \right\} = - \frac{i}{\pi} \left(
\frac{v}{\omega_0} \right) \mbox{sgn}(t_1-t_2)\,
\frac{\theta(v|t_1-t_2|-|{\bf r} - {\bf s}|)}
     {\sqrt{v^2|t_1-t_2|^2-|{\bf r} - {\bf s}|^2}}.
\end{equation}
Carrying out the time-ordered integration over $t_1$ and $t_2$, we
obtain Eq. (\ref{eq:GIsubohmic}).

\subsection{Ohmic case}
\label{sec:imag_ohmic}

For $s=0$, we notice that
\begin{equation}
\frac{1}{2} \left\{ [ f ({\bf r}, t_1), f ({\bf s}, t_2)] + [ f ({\bf
    s}, t_1), f ({\bf r}, t_2)] \right\} = - \frac{i}{\pi} \left(
\frac{v}{\omega_0} \right)^2 \frac{d}{dt_2} \frac{d}{dt_1}
\int_0^\Lambda \frac{dk}{k}\, J_0 (k |{\bf r}-{\bf s}|) \sin[ k v (t_1
  - t_2)].
\end{equation}
Using the integral \cite{gradshteyn_Iohmic}
\begin{equation}
\int_0^\infty \frac{dx}{x}\, J_0(\beta x)\, \sin(\alpha x) =
\mbox{sgn}(\alpha) \left[ \frac{\pi}{2}\, \theta(|\alpha| - |\beta|)
  + \mbox{arcsin} \left( \frac{|\alpha|}{|\beta|} \right)\,
  \theta(|\beta|-|\alpha|) \right]
\end{equation}
we obtain
\begin{eqnarray}
\frac{1}{2} \left\{ [ f ({\bf r}, t_1), f ({\bf s}, t_2)] + [ f ({\bf
    s}, t_1), f ({\bf r}, t_2)] \right\} & = & - \frac{i}{\pi} \left(
\frac{v}{\omega_0} \right)^2 \frac{d}{dt_2} \frac{d}{dt_1} \left[
  \frac{\pi}{2}\, \theta(|t_1-t_2| - |{\bf r} - {\bf s}|)
  \right. \nonumber \\ & & \left. +\ \mbox{arcsin} \left(
  \frac{|t_1-t_2|}{|{\bf r} - {\bf s}|} \right)\, \theta(|{\bf r} -
       {\bf s}|-|t_1-t_2|) \right].
\end{eqnarray}
Carrying out the time-ordered integration in $t_1$ and $t_2$, we
obtain Eq. (\ref{eq:GIohmic}).

\subsection{Super-Ohmic case}
\label{sec:imag_super}

Similarly to the Ohmic case, for $s=1/2$ we write
\begin{equation}
\frac{1}{2} \left\{ [ f ({\bf r}, t_1), f ({\bf s}, t_2)] + [ f ({\bf
    s}, t_1), f ({\bf r}, t_2)] \right\} = - \frac{i}{\pi} \left(
\frac{v}{\omega_0} \right)^3 \frac{d}{dt_2} \frac{d}{dt_1}
\int_0^\Lambda dk\, J_0 (k |{\bf r}-{\bf s}|) \sin[ k v (t_1 - t_2)].
\end{equation}
Using the integral in Eq. (\ref{eq:GR6.761.7}), we obtain
\begin{eqnarray}
\frac{1}{2} \left\{ [ f ({\bf r}, t_1), f ({\bf s}, t_2)] + [ f ({\bf
    s}, t_1), f ({\bf r}, t_2)] \right\} & = & - \frac{i}{\pi} \left(
\frac{v}{\omega_0} \right)^3 \frac{d}{dt_2} \frac{d}{dt_1} \left[
  \frac{\theta(|t_1-t_2| - |{\bf r} - {\bf
      s}|)}{\sqrt{v^2|t_1-t_2|^2-|{\bf r}-{\bf s}|^2}} \right].
\end{eqnarray}
Carrying out the integration the time-ordered integrations on $t_1$
and $t_2$, we arrive at Eq. (\ref{eq:GIsuperohmic}).
\end{widetext}
%

\section{Single-qubit flipping probability}
\label{sec:AppendixD}

We can obtain a compact expression for the single-qubit-bath evolution
operator in Eq. (\ref{eq:Uj}) by following essentially the same steps
shown in Appendix \ref{sec:appendixA}. The only formal difference is
that summations over all qubits in the lattice have to be replaced by
a term corresponding to a single qubit $j$. Thus, considering
Eq. (\ref{eq:evolution}), the result is
\begin{equation}
\label{eq:Ujshort}
U_j(\Delta) = \chi_j\, : \exp \left[ - \frac{i\lambda}{2} F_{{\bf
      r}_j} (\Delta)\, \sigma_j^x \right] :,
\end{equation}
where
\begin{equation}
\chi_j = \exp \left[ -\frac{\lambda^2}{8} {\cal G}^{(R)}_{{\bf
      r}_j{\bf r}_j} (\Delta) \right].
\end{equation}
since ${\cal G}^{(I)}_{{\bf r}_j{\bf r}_j} (\Delta) = 0$. 

Consider now the change of basis
\begin{eqnarray}
|\uparrow_j\rangle & = & \frac{1}{\sqrt{2}} \left( |\uparrow_j\rangle_x
+ |\downarrow_j\rangle_x \right), \\ |\downarrow_j\rangle & = &
\frac{1}{\sqrt{2}} \left( |\uparrow_j\rangle_x - |\downarrow_j\rangle_x
\right),
\end{eqnarray}
which allows one to write
\begin{eqnarray}
\langle \uparrow_j| U_j(\Delta) | \downarrow_j \rangle & = &
\frac{1}{2} \left\{ \exp \left[ - \frac{i\lambda}{2} F_{{\bf
      r}_j}(\Delta) \right] \right. \nonumber \\ & & \left. -\ \exp
\left[ \frac{i\lambda}{2} F_{{\bf r}_j}(\Delta) \right] \right\}
\end{eqnarray}
since $:e^{- \frac{i\lambda}{2} F_{{\bf r}_j}(\Delta)}:\ =
\chi_j^{-1}\, e^{- \frac{i\lambda}{2} F_{{\bf r}_j}(\Delta)}$ [see
  Eqs. (\ref{eq:normalorder}) and (\ref{eq:vv})]. We can then
write
\begin{widetext}
\begin{eqnarray}
\langle \uparrow_j | U^\dagger_j(\Delta) |\downarrow_j \rangle \langle
\downarrow_j | U_j(\Delta) |\uparrow_j \rangle & = & \frac{1}{4}
\left\{ 2 - \exp \left[ i\lambda\, F_{{\bf r}_j}(\Delta) \right] -
\exp \left[ - i\lambda\, F_{{\bf r}_j}(\Delta) \right] \right\}
\nonumber \\ & = & \frac{1}{4} \left\{ 2 - \chi_j^2 :\exp \left[
  i\lambda\, F_{{\bf r}_j}(\Delta) \right]: - \chi_j^2 :\exp \left[ -
  i\lambda\, F_{{\bf r}_j}(\Delta) \right]: \right\},
\end{eqnarray}
\end{widetext}
which yields Eq. (\ref{eq:probexp}) when the expectation value over
the bath vacuum is taken. Now we can insert Eq. (\ref{eq:Ujshort})
into Eq. (\ref{eq:prob}) to obtain Eq. (\ref{eq:probexp}).
%



\begin{thebibliography}{35}
\bibitem{footnote2} When the syndromes point to an error, one can try
  to recover the state by flipping back qubits. However, there is no
  guarantee that the appropriate string of qubits is being flipped. As
  a result, the recovery operation may induce a logical error.
\bibitem{gradshteyn_Rsubohmic} This expression is obtained using
  integration by parts and Eqs. 6.56.1.14, 6.693.2, and 6.693.1 from
  Ref. \onlinecite{gradshteyn}.
\end{thebibliography}
\end{document}